\documentclass[final,authoryear,5p,times,twocolumn]{elsarticle}

\usepackage{amssymb}
\usepackage{epsfig}
 \usepackage{subfigure}
\usepackage{graphicx}
\def\astrobj#1{#1}

\journal{New Astronomy}

\begin{document}

\begin{frontmatter}



\title{Two-site CCD observations and spectroscopy of  \astrobj{HD~207331}: a new $\delta$ Scuti variable
in Cygnus}


\author[oan]{L. Fox Machado\corref{cor}}
\ead{lfox@astrosen.unam.mx}

\author[oan]{W.J. Schuster}

\author[iac]{C. Zurita}

\author[oan]{J.~S. Silva}

\author[oan]{R. Michel}

\cortext[cor]{Corresponding author. Tel.: +52 6461744580; fax +52
6461744607 }

\address[oan]{Observatorio Astron\'omico Nacional, Instituto de
Astronom\'{\i}a -- Universidad Nacional Aut\'onoma de M\'exico, Ap.
P. 877, Ensenada, BC 22860, Mexico}

\address[iac]{Instituto de Astrof\'{\i}sica de Canarias
C/V\'{\i}a L\'actea s/n, 38205 La Laguna, Tenerife, Spain}

\begin{abstract}
 The results of an  observational campaign on the new $\delta$ Scuti
  pulsator \astrobj{HD~207331} are reported. The star was observed
photometrically from August 26 to September 2, 2009 from the
Observatorio San Pedro M\'artir (0.84-m telescope, Mexico) and the
Observatorio del Teide (0.80-m telescope, Spain).  An overall run of
53.8 h of useful data was collected from the two sites. Four
oscillation frequencies for \astrobj{HD~207331} have been found
above a 99\% confidence level. These results confirm the
multiperiodic pulsation nature of the star suggested in previous
observations with sparse data. Spectroscopic observations carried
out in 2009 allowed us to derive its spectral type and luminosity
class  as well as to estimate its rotation rate. A simple comparison
with models is performed.

\end{abstract}


\end{frontmatter}


\section{Introduction}
\label{sec:int}

The $\delta$ Scuti-type pulsators  are stars with masses between 1.5
and 2.5 $M_{\odot}$ located at the intersection of the classical
Cepheid instability strip with the main sequence. They have spectral
types A and F, a period range between 0.5 h and 6 h, and generally
pulsate with a large number of radial and nonradial modes excited by
the $\kappa$-mechanism associated with the second Helium ionization
zone. These oscillation modes penetrate to different depths inside
the star.  Thus, $\delta$ Scuti stars provide a good opportunity to
probe the internal structure of intermediate mass stars.

Since most of the $\delta$ Scuti stars are short period variables
with typical photometric amplitude of 20 mmag, their oscillations
can be  easily detected from the ground.  In fact, several $\delta$
Scuti stars have been discovered accidentally when taken as
reference stars for observations of well known $\delta$ Scuti stars
[e.g. \citet{fox1,fox2}; \citet{li}]. However to resolve their rich
oscillation spectrum from the ground, high quality long time series
are required which can only be obtained by means of observations
from different sites distributed in longitude around the Earth (e.g.
\citealt{li1,costa}). Therefore, to characterize the pulsation
spectrum  of a  new $\delta$ Scuti star additional observational
efforts are needed.

In this paper we present the results of photometric and
spectroscopic follow-up observations of \astrobj{HD~207331}, a new
$\delta$ Scuti star recently discovered in Cygnus, aimed at
characterizing its pulsation behavior more accurately.

\section{The object \astrobj{HD~207331}}
\label{sec:obj}

The star \astrobj{HD~207331} ($=$ SAO 51294, BD$+$42 4207, HIP
107557) was discovered to be a new $\delta$ Scuti star by
\cite{schuster}
 when observing  a sample of
A-type stars on the night of September 27, 2007,
 in order to test the six-channel
$uvby$-$\beta$ spectrophotometer attached to the H.~L. Johnson 1.5-m
telescope of the San Pedro M\'artir Observatory, Baja California,
Mexico. The variability of the star was clear, despite the fact that
it was observed less than 2 h. Six hours of CCD photometric
observations on the night of September 30, 2007, with the 0.84-m
telescope of the same observatory confirmed its variability
\citep{schuster}.

 Differential and standard Str\"omgren $(uvby)$ photometry
 of \astrobj{HD~207331} was obtained by  \cite{fox3} using the
 same equipment as  the former
 observations.  In particular, the star was monitored  for about 11 h between
November 11 and November 19, 2007.
    As a result, \cite{fox3} found evidence of at least
    two close frequencies which might explain the resulting beating behavior
    of the light curve.  The derived standard photometry is the
following:  ($V$, $b$--$y$, $m_1$, $c_1$) = (8.329, 0.125, 0.150,
1.018). \cite{fox3}  also estimated a stellar reddening  from the
reddening maps of Schlegel et al. (1998) via the web-page calculator
of the NED (NASA/IPAC Extragalactic Database). The derived intrinsic
colours of \astrobj{HD~207331} are the followings ($V_0$,
$(b$--$y)_0$, $m_0$, $c_0$) = (7.980, 0.044, 0.174, 1.002), and
$\beta = 2.854$, which yield a $M_V = 1.058$ mag, and $d = 242$ pc.

\begin{figure}[!t]
\includegraphics[width=6.0cm,height=6.0cm]{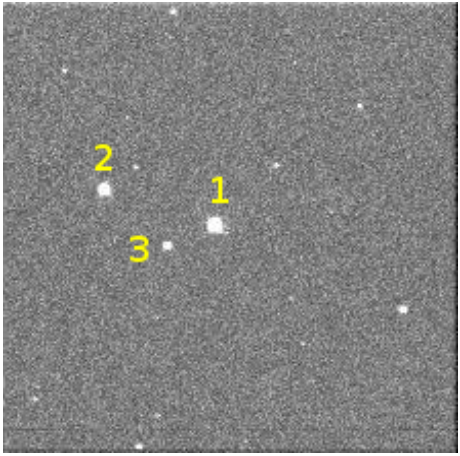}
\caption{CCD imagen of the \astrobj{HD~207331} FOV at the 0.84m
telescope. 1 stands for \astrobj{HD~207331},  2 for the comparison
star, and 3 for the check star. Some properties of the stars are
listed in Table 1. North is up and East is left.} \label{fig:field}
\end{figure}

 The Hipparcos catalogue \citep{perryman}, on the other hand,
 provides a parallax of 3.31 $\pm$ 0.88
mas, from which a distance value of 302 $\pm$ 81 pc can be
estimated. This corresponds to a distance modulus of 7.40 $\pm$ 0.60
mag, which indicates a value of $M_{v}=1.0 \pm$ 0.6 mag for
\astrobj{HD~207331}. Therefore the Hipparcos
 distance is in agreement within 1-$\sigma$ error with
distance reported by \cite{fox3}.
  The Hipparcos catalogue also lists for \astrobj{HD~207331} a
magnitude of $H_{p}$ of 8.3970 $\pm$ 0.0022 mag (median error),
$\pm$ 0.019 mag (scatter) and a $V_{T }$ of 8.335 $\pm$ 0.009 mag
(standard error).

Concerning the spectral classification of \astrobj{HD~207331}, the
SAO Star Catalog J2000 (SAO Staff 1966; USNO, ADC 1990) lists A0.

\begin{figure*}[!t]
\begin{center}
\includegraphics[width=14cm,height=22cm]{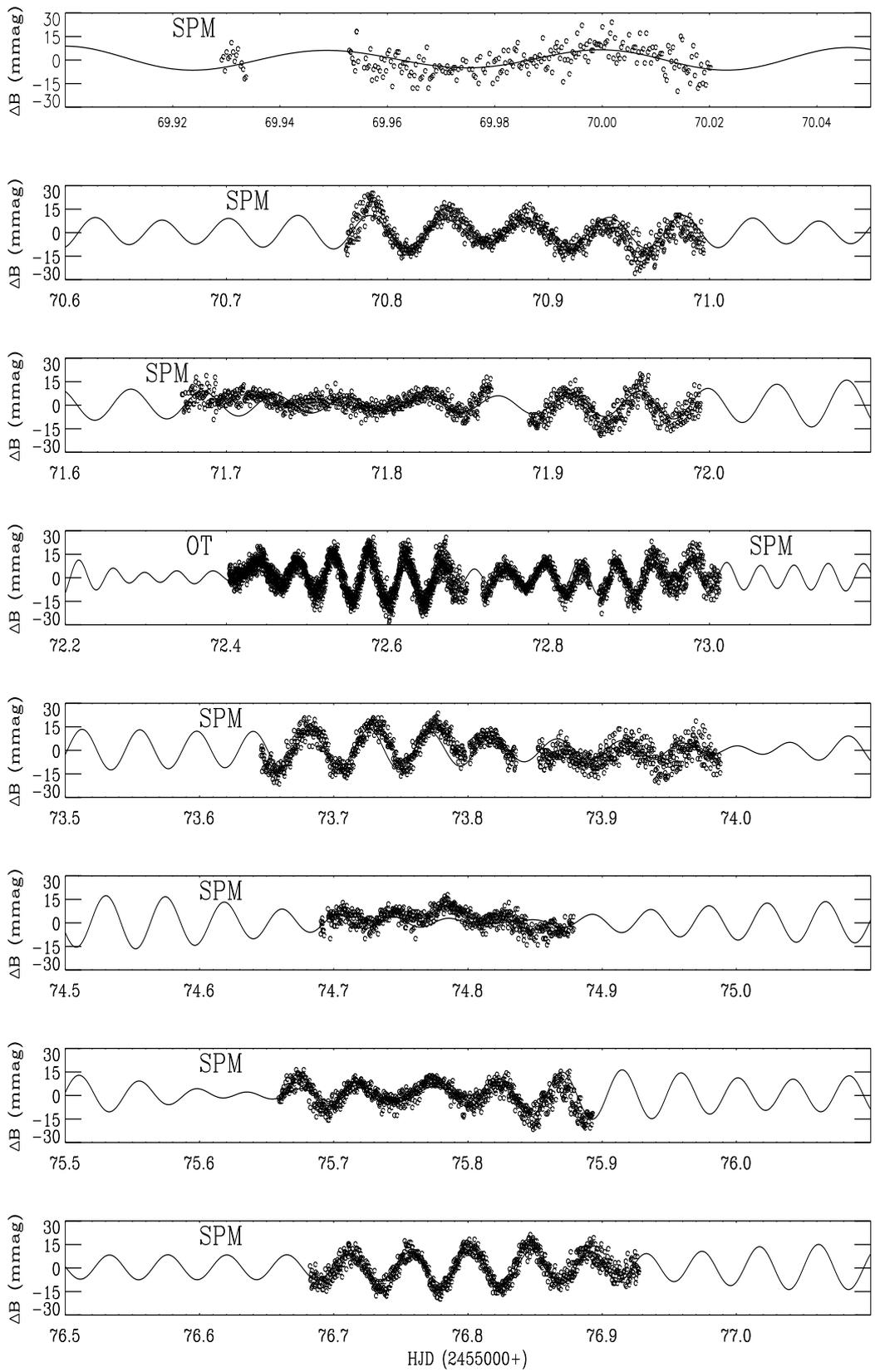}
\caption{Differential light curves of \astrobj{HD~207331}. The name
of the observatory involved is indicated in each panel.}
\label{fig:curves}
\end{center}
\end{figure*}

\section{Observations and data reduction}
\label{sec:obs}

\subsection{CCD photometric observations}

The two observatories  involved in the observational campaign are
the following:\\

{ -Observatorio del Teide (OT, Spain):} the data were collected
using the IAC 80 telescope (0.80-m). The images were acquired with
CAMELOT,
a CCD detector of $2048 \times 2048$ pixels. Observer: CZ.\\
 {-Observatorio de San Pedro M\'artir (SPM, Mexico):} the observations were
 performed using the 0.84-m telescope.
 The CCD was a SITe1 of $1024
\times 1024$ pixels. Observers: WJS, JS, LFM. Observer's
abbreviations correspond to the initials of the
co-authors.\\

Table~\ref{tab:log} gives the log of observations. Bad weather
conditions at the OT did not allow us to get more than one night of
data. However, a total amount of 53.8 h of useful data was obtained
from the two sites.

\begin{table*}
\caption{Log of observations.  Observing time is expressed in
hours.}
\begin{tabular}{cccccr}

\hline

    Day& UT Date 2009 &  Start Time  & End Time &  OT  & SPM$\;$ \\
      &                &(HJD 2455000+)&(HJD 2455000+)&&       \\
 \hline
      1&    Aug 26    &    69.93     &   70.02  &  -   & 2.185 \\
      2&    Aug 27    &    70.77     &   71.00  &  -   & 5.335 \\
      2&    Aug 28    &    71.67     &   72.70  &7.086 & 7.731 \\
      4&    Aug 29    &    72.72     &   73.01  &  -   & 7.076 \\
      5&    Aug 30    &    73.65     &   73.99  &  -   & 8.218 \\
      6&    Aug 31    &    74.69     &   74.88  &  -   & 4.536 \\
      7&    Sep 01    &    75.66     &   75.89  &  -   & 4.592 \\
      8&    Sep 02    &    76.68     &   76.93  &  -   & 7.085 \\
    \hline
       &    Begin     &     End      &Total Time&  OT  & SPM$\;$ \\
       &    Aug 26    &    Sep 02    & 53.844   &7.086 &46.758 \\
\hline
\end{tabular}
\label{tab:log}
\end{table*}

\begin{table*}[!t]\centering
  \setlength{\tabcolsep}{1.0\tabcolsep}
 \caption{Position, magnitude, and spectral type of target, comparison,
and check stars observed in the CCD frame.}
  \begin{tabular}{lccccc}
\hline
Star       &       ID       &    RA     &   Dec      &   V   & SpTyp \\
           &                &  (2000.0) & (2000.0)   & (mag) &       \\
\hline
Target     & \astrobj{HD~207331}    & 21 47 02  &$+$43 19 19 &  8.3  & $A0$  \\
Comparison & \astrobj{BD$+$42 4208}  & 21 47 12  &$+$43 19 51 &  9.4  & $A0$  \\
Check      & \astrobj{TYC 3196-1243-1}& 21 47 06  &$+$43 18 58 & 10.9  &  -    \\
\hline
\end{tabular}
\label{tab:stars}
\end{table*}

\medskip
The observations were obtained through a Johnson $B$ filter.
Figure~\ref{fig:field} shows a typical image of the CCD's field of
view ($6.8.' \times 6.8'$) at the 84-cm telescope of the San Pedro
M\'artir observatory. The target star is labeled with number 1;
comparison and check stars with 2 and 3, respectively. As can be
seen this is an uncrowded star field. Although the CCD's FOV at the
IAC-80 telesope of the Teide observatory is larger ($\sim$ $11'
\times 11'$), no other suitable comparison stars could be observed.
 Table~\ref{tab:stars} shows the main observational parameters
corresponding to the target and comparison stars as taken from the
SIMBAD database operated by the CDS (Centre de Donn\'ees
astronomique de Strasbourg).

\medskip
 Sky flats,
dark and bias exposures were taken every  night at both sites. All
data were calibrated and reduced using standard IRAF routines.
Aperture photometry was implemented to extract the instrumental
magnitudes of the stars. The differential magnitudes were normalized
by subtracting the mean of differential magnitudes for each night.
In Figure~\ref{fig:curves} the entire light curves,
\astrobj{HD~207331} - Comparison, are presented. As can be seen from
the fourth panel (from top to bottom) no overlapping of the
observations was obtained for the one night of observing at OT, the
28th of August 2009.

\begin{figure}[!t]
\includegraphics[width=8cm]{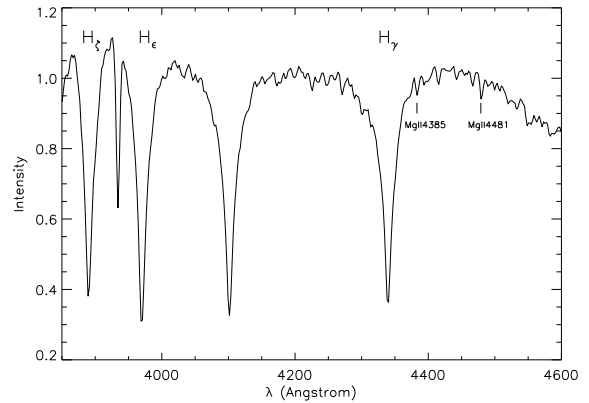}
  \caption{Normalized spectrum of \astrobj{HD~207331}. For sake of clarity only the
  region between $3800\, \AA - 4600\, \AA$ is shown.}
  \label{fig:spectra}
\end{figure}

\subsection{Spectroscopic observations}\label{sec:spec_obs}
Spectroscopic observations were conducted at the 2.12-m telescope of
the SPM observatory during  September 27, 2009 (UT). We used the
Boller \& Chivens spectrograph installed in the Cassegrain focus of
the telescope. The 600 lines/mm gratting was used to cover a
wavelength range from 3900 to 6000 \AA. A
  dispersion of 2.05 \AA\, per
pixels with a resolution of 5.0  \AA\, was employed. The SITe3 $1024
\times 1024$ pixel CCD with a 0.24 $\mu$m pixel size was attached to
the spectrograph. The spectra were reduced in the standard way using
the IRAF package. Fig.~\ref{fig:spectra} shows a close-up of the
normalized spectrum at region $3800\, \AA - 4600\, \AA$.
 The spectral type was derived by comparing the
normalized spectra with those of well classified stars available in
the literature. An additional check of the spectral type was done
from the ratio of lines Mg II 4481 and Mg II 4385.
  We have derived a spectral type of A0V for
\astrobj{HD~207331}.

The projected rotational velocity of the star ($v\,sin\,i$) can be
estimated from the empirical calibration $v\,sin\,i$ vs FWHM given
by \cite{bush} who determined the $v\,sin\,i$ of 118 $\delta$ Scuti
variables. To do so, we have measured the FWHM for the 4501 $\AA$ Ti
II  and 4508 $\AA$ Fe II lines.  Then, we have applied the
relationship \citep{bush}:

\begin{equation}
v\,sin\,i= -33.7FWHM^{2} +  226.0FWHM -215.3
\end{equation}

\noindent As a result a $v\,sin\,i = 133 \pm 10$ km/s was derived
for \astrobj{HD~207331}.

In order to test the validity of the calibration for our telescope
system we also have derived the $v\,sin\,i$ values for $\delta$
Scuti variables \astrobj{7~Aql} and \astrobj{8~Aql} whose spectra
were recorded by \cite{fox5} with the same telescope, equipment and
configuration as ours.  These stars have accurately determined
$v\,sin\,i$ in the literature. We have found very good agreement
between the derived projected rotational velocities from equation
(1) and those listed in the literature for \astrobj{7~Aql} and
\astrobj{8~Aql}.

\section{Period analysis}

The period analysis has been performed by means of standard Fourier
analysis and least-squares fitting. In particular, the amplitude
spectra of the differential time series were obtained by means of
Period04 package \citep{lenz}, which utilizes Fourier as well as
multiple least-squares algorithms. This computer package allows us
to fit all the frequencies simultaneously in the magnitude domain.

\begin{figure}
\includegraphics[width=8.0cm,height=10.0cm]{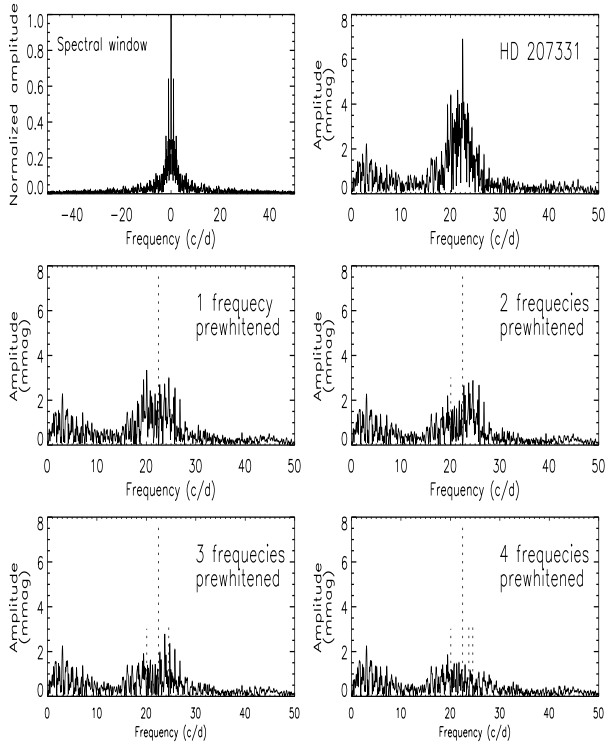}
\caption{Spectral window (first plot). Pre-whitening process in
\astrobj{HD~207331} (from second plot). In each plot, from left to
right, the highest amplitude peak is selected and removed from the
time series, and a new spectrum is obtained.} \label{fig:spec}
\end{figure}

\medskip
The spectral window in amplitude of the observations is shown in the
first plot of Fig.~\ref{fig:spec}. The amplitude spectrum of the
differential light curve, HD 207331 - Comparison is depicted in the
next plot.
  The subsequent plots in the figure, from left to right,
illustrate the prewhitening process of the frequency peaks in each
amplitude spectrum.

\medskip
The frequencies have been extracted by means of a standard
prewhitening method. In order to decide which of the detected peaks
in the amplitude spectrum can be regarded as intrinsic to the star,
Breger's criterion has been followed \citep{breger},  where it was
shown that the signal-to-noise ratio (in amplitude) should be at
least 4 in order to ensure that the extracted frequency is
significant.

\medskip
The frequencies, amplitudes, and phases  are listed in
Table~\ref{tab:frec}. Four significant frequencies have been
detected in \astrobj{HD~207331}. A comparison of this four-frequency
solution to the data is displayed in Fig.~\ref{fig:curves} with the
solid line.

\begin{table}[!t]\centering
  \setlength{\tabcolsep}{1.0\tabcolsep}
 \caption{Frequency peaks detected in the light curve:  \astrobj{HD~207331} - Comparison.
S/N is the signal-to-noise ratio in amplitude after the prewhitening
process.} \label{tab:frec}
  \begin{tabular}{ccccccc}
\hline
    &Freq. &Freq.  & Period&  A   & $\varphi$/($2\pi$)& $S/N$ \\
    &(c/d) &($\mu$Hz )&hours &(mmag)&                   &       \\
\hline
  $f_1$& 22.4880 &260.186&0.937& 7.76 &      0.18        &  13.7 \\
  $f_2$ &20.0923 &232.468&0.837& 3.02 &       0.90        &   5.1 \\
  $f_3$& 24.5384 &283.909&1.022& 3.22 &       0.32        &   6.3 \\
  $f_4$& 23.7409 &274.682&0.989& 3.18 &       0.85        &   5.9 \\
 \hline
\end{tabular}
\end{table}

\section{Discussion }

From the differential light curves it can be noted that the beating
 behavior seen in the short time series of \cite{fox3} is present.  This is a common
 characteristic of the
light curves of $\delta$ Scuti stars. The amplitude spectrum shows a
concentration of high signal-to-noise peaks between 18--28 cycle
day$^{-1}$. In particular, the highest amplitude peak is located at
22.49 cd$^{-1}$ (260.19 $\mu$Hz), and the next significant frequency
is located at 20.09 cd$^{-1}$ (232.47 $\mu$Hz ).

\medskip
 \cite{fox3} detected a significant frequency peak at $\sim$ 21.1
cd$^{-1}$ (244.1 $\mu$Hz) with an amplitude of 6 mmag in Str\"omgren
$y$ band.  As can be seen in Table~\ref{tab:frec}, this is well
within the period of the modes detected in this campaign, taking
into account that the complex window function of those observations
does not allow a precise comparison. Nonetheless, this peak is most
likely a side-lobe of the second frequency $(f_2)$ detected in our
campaign.
 No equally spaced close frequency pairs were found in the frequency
pattern of \astrobj{HD~207331}.

\subsection{Preliminary comparison with theoretical
models}\label{sec:models}

\begin{figure}[!t]
\centering
\includegraphics[width=8cm]{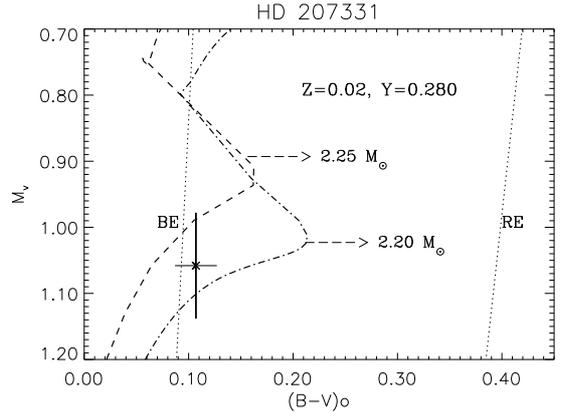}
\caption{Position of \astrobj{HD~207331} in the HR diagram.
Evolutive sequences of non-rotating models without overshooting are
shown by dashed $(M/M_{\odot}=2.25)$ and dot-dashed lines
$(M/M_{\odot}=2.20)$. The borders of the instability strip are shown
by dotted lines.} \label{fig:models}
\end{figure}

Figure~\ref{fig:models} shows the observed position of
\astrobj{HD~207331} (asterisk) in the HR diagram and its associated
uncertainty (cross upon the asterisk). Since the Hipparcos distance
for \astrobj{HD~207331} has a large relative error, we have
considered in Fig.~\ref{fig:models} the St\"romgren absolute
magnitude derived by \cite{fox3}, namely  $M_V = 1.058$. Error bars
of 0.1 mag for $M_{V}$ and 0.02 mag for $(B-V)$ have been adopted.

\medskip
  Two main
sequence evolutionary tracks that approximately match the observed
position of  \astrobj{HD~207331} are shown by dashed and dot-dashed
lines. These evolutionary sequences were computed as explained in
\cite{fox4} by using the CESAM evolution code \citep{morel} with
input physics appropriate to $\delta$ Scuti stars and a chemical
initial composition of $Z=0.02$ and $Y=0.28$. Also shown are the
observed instability strip boundaries from Rodr\'{\i}guez et al.
(1994). According to the models depicted in Fig.~\ref{fig:models},
for a solar metallicity, HD 207331 would have a mass between 2.20
$M_{\odot}$ and 2.25 $M_{\odot}$. The  age of HD 207331
corresponding to these evolutionary models is $\sim$ 527 Myr and
$\sim$ 513 Myr, respectively.

\medskip
We now use these models to derive possible radial and non-radial
modes in order to obtain some insights into the pulsation behaviour
of \astrobj{HD~207331}. The adiabatic eigenfrequencies were computed
using the code FILOU \citep{suarez}.  The following identifications
of the modes are possible, but not unique: $f_{1}$ -- ($l=0$,
$n=5$), $f_{2}$ -- ($l=2$, $n=2$), $f_{3}$ -- ($l=1$, $n=4$),
$f_{4}$ -- ($l=2$, $n=3$). We note that since these identifications
were computed from just few equilibrium models, they should be
considered as preliminary. Moreover, we have neglected the
rotational effects in spite the fact that \astrobj{HD~207331} is
most likely a high rotating star. In fact, the rotation effects are
important not only in the position of the star in a HR diagram
\citep{perez} but also in the computation of oscillation frequencies
\citep{suarez1}.

\medskip
 Nevertheless, these simple computations indicates that \astrobj{HD~207331}
may pulsate in radial and non-radial modes typical among $\delta$
Scuti stars. Since the star is located near blue edge of instability
strip, it could be pulsating in higher overtones than the
fundamental.

\section{Conclusions}

A summary of the follow-up observations which led to the
characterization of the new $\delta$ Scuti star HD 207331 has been
presented. The star was observed from two observatories distributed
in longitude around the Earth.
  An overall run of 53.8 h
of useful data was collected from the two sites on nine observing
nights. A period analysis reveals that \astrobj{HD~207331} is a
multiperiodic pulsating star with at least four oscillation modes.

\medskip
A simple comparison with theoretical modes has been performed. The
star shows complicated pulsations as do most of the $\delta$ Scuti
stars. \astrobj{HD~207331} seems to pulsate with low-order-$p$ modes
typical among $\delta$ Scuti stars.

\medskip
 The analysis of a few low
resolution spectra  points to it being a fast rotating $\delta$
Scuti star of spectral type A0V.

\medskip
 To date our observations represent the most
extensive work on HD 207331.

\bigskip
{\bf \noindent Acknowledgements}

This work has received financial support from the UNAM via PAPIIT
grant IN114309. WJS acknowledges financial support from CONACyT by
way of grant 49434-F.  Based on observations collected at the 0.84 m
telescope at the Observatorio Astron\'omico Nacional at San Pedro
M\'artir, Baja California, Mexico, and at the IAC-80 telescope
operated by the Instituto de Astrof\'{\i}sica de Canarias in the
Spanish Observatorio del Teide. Special thanks are given to the
technical staff and night assistants of the San Pedro M\'artir and
Teide Observatories. This research has made use of the SIMBAD
database operated at the CDS, Strasbourg (France).

\bigskip
{\noindent \bf References}
\medskip

\bibliographystyle{elsarticle-harv}

\end{document}